\documentclass[10pt,journal]{IEEEtran}
\newtheorem{definition}{Definition}
\usepackage{graphicx}
\usepackage{algpseudocode}
\usepackage{algorithm}
\usepackage{amsmath}
\usepackage{caption}
\usepackage{subcaption}
\usepackage{cite}
\usepackage{xcolor}

\makeatletter
\renewcommand{\ALG@beginalgorithmic}{\small}
\makeatother

\begin{document}

\title{Application-aware Retiming of Accelerators:\\ A High-level Data-driven Approach}

\author{Ana~Lava,~\IEEEmembership{Member,~IEEE,}
        Mahdi~Jelodari~Mamaghani,~\IEEEmembership{Member,~IEEE,}
        Siamak~Mohammadi,~\IEEEmembership{Member,~IEEE,}
        and~Steve~Furber,~\IEEEmembership{Fellow,~IEEE}
%\IEEEcompsocitemizethanks{
%\IEEEcompsocthanksitem M. Jelodari Mamaghani is with the University of Southern California, USA as an exchange scholar.\protect\\ E-mail: see http://apt.cs.manchester.ac.uk/people/mamagham/
%\IEEEcompsocthanksitem S. Furber are with the University of Manchester, Oxford Road, Manchester, UK, M13 9PL.\protect\\ Email: Steve.Furber@manchester.ac.uk
%\IEEEcompsocthanksitem A. Lava and S. Mohammadi are with University of Tehran, Fanni, Tehran, Iran.}\thanks{Manuscript received October 3, 2016.}
}

\markboth{IEEE Design \& Test Journal Special Issue on Special Issue on Hardware Accelerators for Data Centers}%
{Shell \MakeLowercase{\textit{et al.}}: Bare Advanced Demo of IEEEtran.cls for IEEE Computer Society Journals}

\IEEEtitleabstractindextext{%
\begin{abstract}

Flexibility at hardware level is the main driving force behind adaptive systems whose aim is to realise microarhitecture deconfiguration `online'. This feature allows the software/hardware stack to tolerate drastic changes of the workload in data centres. With emerge of FPGA reconfigurablity this technology is becoming a mainstream computing paradigm. Adaptivity is usually accompanied by the high-level tools to facilitate multi-dimensional space exploration. An essential aspect in this space is memory orchestration where on-chip and off-chip memory distribution significantly influences the architecture in coping with the critical spatial and timing constraints, e.g. Place \& Route. 
   
This paper proposes a memory smart technique for a particular class of adaptive systems: \textit{Elastic Circuits} which enjoy slack elasticity at fine level of granularity. We explore retiming of a set of popular benchmarks via investigating the memory distribution within and among accelerators. The area, performance and power patterns are adopted by our high-level synthesis framework, with respect to the behaviour of the input descriptions, to improve the quality of the synthesised elastic circuits.

\end{abstract}

}

\maketitle

\IEEEdisplaynontitleabstractindextext

\IEEEpeerreviewmaketitle

\ifCLASSOPTIONcompsoc
\IEEEraisesectionheading{\section{Introduction}}
\else
\section{Introduction}

\fi

\IEEEPARstart{E}{lasticity} has emerged as a property that implies flexibility in adapting resources, communication or timing in different areas of computer architecture.  Elasticity in cloud computing is usually devoted to the resource management capability by the cloud services which are tolerant against changes in the user inquiries; On the other hand it is used by neuroscientists to refer to \textit{plasticity} in the brain which leads to dynamic structural changes between neural clusters when learning a new skill~\cite{Morrison2008}. Elasticity in digital circuits follows the same concept and particularly is referred to the flexibility against environmental dynamics such as temperature or process variation which may violate timing of the functional units in a circuit~\cite{petrica2013flicker}.

There are always costs associated with implementing elasticity which may vary based on several factors. For instance, in cloud computing an advanced protocol is necessary to determine the user response times and, accordingly, reschedule the routing and resource allocation. Similarly, in digital circuits handshake protocols are leveraged to realise elasticity which requires special control circuitry to realise the interlocks between stages. Additional circuitry may impose a prohibitive impact on power and performance. Therefore elasticity has to be inserted with careful consideration to minimise the unnecessary costs. De-Elastisation~\cite{mamaghani2015elastisation} is proposed to remove elasticity `selectively' from the design by establishing rigid synchronous blocks which may run at different frequencies. By forming coarse synchronous islands the available fine grained adaptivity is sacrificed. Therefore a clustering approach would be necessary to trade off the cost of adaptivity at different levels of granularity.\\ 

Elastic dataflow circuits have been proposed to address the performance overhead of control-driven circuits through concurrency and meanwhile offer a distributed control mechanism which scales effectively and allows local optimisations, such as re-synthesis, to take place. The channel-based~(aka CSP-based) class of these circuits enjoys a property known as \textit{slack elasticity} which allows retiming and recycling of these circuits without breaking their functionality. In other words any degree of pipelining, including compile-time transformations, can be explored in this regard. At system level, varying the available slack gives us the chance to look at different flavours of memories and consider transforming them to each other to exploit the existing potentials. The concept of smart memories and adaptive memory-based architectures have been studied in depth~\cite{firoozshahian2009smart}.

\begin{figure}[!t]
\centering
\includegraphics[totalheight=5.5cm, width=0.5\textwidth]{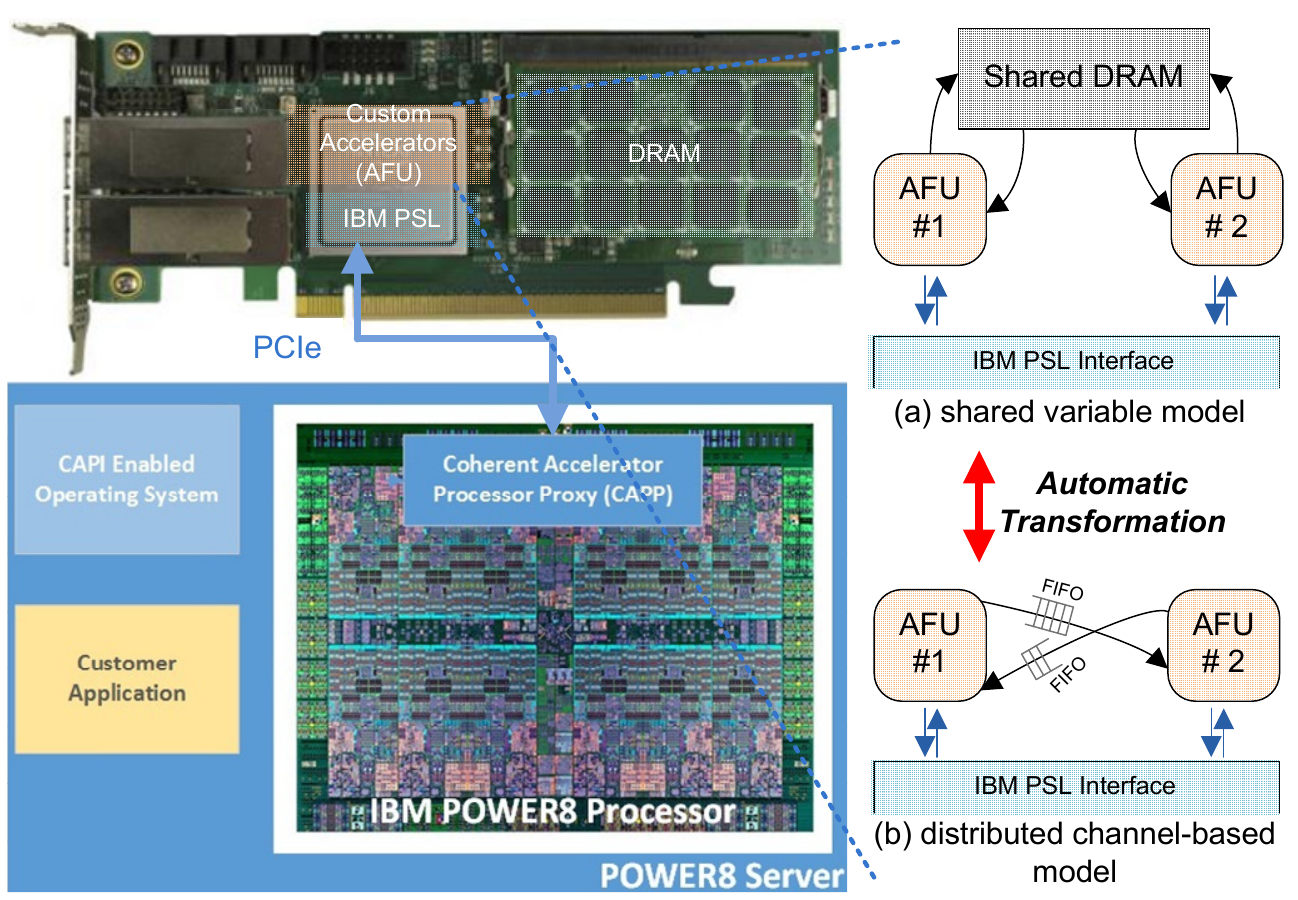}
\caption{Memory type~(shared, communication, storage) distribution across a system are transformable based on our proposed channel based elastic model for accelerators. }
\label{fig:system}

\vspace{-0.6cm}
\end{figure}

%\begin{figure}[!t]
%\centering
%\includegraphics[totalheight=5cm, width=0.5\textwidth]{retiming.pdf}
%\caption{Automatic transformation of a shared memory type~(variable) to communication type~(1-depth FIFOs) via retiming: results in a pipelined architecture with a three-fold performance gain.}
%\vspace{-20pt}
%\label{fig:retiming}
%\end{figure}

In this paper we exploit slack elasticity to refine the system architecture automatically with regard to the energy budget or performance constraints. Memory types are considered interchangeably via retiming without affecting the functionality (Figure~\ref{fig:system}). Unlike \textit{Dynamic Voltage and Frequency Scaling} our proposed technique is applicable at a fine level of granularity, therefore it is able to handle the power/performance trade-off more effectively.\\
%Elastic circuits both in the asynchronous and synchronous domains are exercised in this work. Our results demonstrate that in an un-pipelinable dataflow application such as GCD, using the shared memory type rather than communication FIFOs (or buffers) may result in a higher throughput than other extreme approaches meanwhile yielding better area and power. However this increases the dynamic power as the throughput goes up. The clock frequency could be set to a desired amount equal or less than the operational frequency (when time borrowing is enabled). In less pipeline friendly cases by over-buffering a slight improvement over the operational clock is expectable. The framework permits consideration of trade-offs with respect to constraints set by the designer.

The organization of the paper is as follows: Section~2 presents our high-level dataflow architectural model and following that Section~3 formulates the design constraints based on our high-level CSP model. Section~4 proposes our methodology including a motivation example and our retiming technique. Afterwards, experimental results are given in Section~5. Finally, conclusions and future work ahead are drawn in Section~6.

\section{Architectural Model: Elastic Dataflows}

eTeak~\cite{jelodari2013eTeak} is a dataflow synthesis framework whose features can be grouped into communication and computation domains. From the communication perspective eTeak networks are synthesised in a syntax-directed compilation manner from a CSP-like language called Balsa. Point-to-point communication between computation blocks at hardware level is formed by the primitives of the language, including channels and processes. The networks are `slack elastic' which means the communication channels are capable of storing any degree of tokens.

A `token' is a high-level abstraction of data and control signal. This feature enables modification at the level of pipelining over the channels without affecting the behaviour of the circuit.  From the computational perspective the network is built based on the macro-module style~\cite{MACROMODULE} with separate `go' and `done' activation/termination signals. These modules are chained in sequence or parallel according to source level directives. The macro-module architecture contributes to a distributed control mechanism where the datapath and the corresponding control are enclosed within a macro-module. Accordingly, modules are controlled locally through handshaking, thus, whenever data become available, computation can start.  This concept has already been introduced in dataflow systems.  Based on this, data-dependent computation becomes possible which means that independent data streaming can exist within a module, which can significantly influence the performance of the circuit.  In addition, it allows the tool to perform functional decomposition over a module and define new boundaries.\\

Explicit buffering is needed to decouple one component from another and to introduce the desired degree of token storage to enable the circuit to function.  The buffering can also allow more transforming synthesis methods (e.g. retiming) to increase circuit parallelism. 

%\subsection{eTeak Component Set}

eTeak primitives are the same as any regular dataflow architecture. The following describes each briefly:\\ 

%\begin{figure}[!t]
%\centering
%\includegraphics[totalheight=5cm, width=0.45\textwidth]{eTeak-primitives}
%\caption{The eTeak primitive components used to implement dataflow circuits.}
%\label{fig:teakComp}
%\end{figure}

\begin{itemize}
\item Steer~(S) -- Chooses an output path based on the input control value attached to data.  Steers are inferred wherever an if/else or case statement is used.  Each parametrised output independently matches the conditions of input and it acts like a data-dependent de-multiplexer.

\item Fork~(F) -- A parameterisable component which can carry any number of bits from input to outputs.  It brings concurrency to the circuit by activating two or more macro-modules at the same time or supplying them with data.

\item Merge~(M) -- Input on one of the input ports is multiplexed towards the output based on first-come-first-served policy; thus, the inputs must be mutually exclusive.  Merge is also parameterisable, which means that it can function as a data or control multiplexer.

\item Join~(J) -- Synchronizes and concatenates data inputs.  A two-way join of n and 0 bits can be used as a conjunction of data and control.

\item Variable~(V) -- Permanent data storage.  A variable in the eTeak dataflow network has a single write port and multiple, parameterisable read ports.  The reads and writes are distinguished and put into separate stages.  Variables allow complicated control activity without incurring the cost of always moving data along with control around a circuit.  `wg/wd' and `rg/rd' (go/done) pairs make all writes data initiated and control token completed, all reads control token initiated and data delivery terminated.  The variable can be considered as a multi-bit register in which a read from it means assigning the contents of the register to the output wire.  Similarly a write to a variable could be translated as assigning the current value of the input wire to the register.

\item Operator~(O) -- The only components which can manipulate data.  Inputs
are formed into a single word.  All data transforming operations are
done within this component, including verifying a condition or other
operations.

\item Initial~(I) -- This component holds and initial value and can insert values such as activation into the network. When a top-level module is generated to start over and over~(within the loop structure) a `go' signal may not exist. In that case I initializes the activation at each round.

\item Buffer~(B) -- Data storage and channel decoupling.  Buffers provide storage for valid and empty tokens and they are the only components that initiate and take an active part in handshaking; all other components are ``transparent'' to the handshaking.  A buffer may input and store a new token (valid or empty) from its predecessor if its successor buffer has input and stored the token which it was previously holding %~(Figure~\ref{fig:Time-Borrowing}).

\item Arbiter~(A) -- It takes a number of input channels, and gives a single output channel, forwarding on any data from input channels to the output channel, fairly choosing between concurrent accesses. This component could be used as a memory or a bus arbiter to control several master accesses. If the masters are clocked at the same speed then arbiters could be implemented as synchronous arbiters~(TeakM) otherwise they should be realised as asynchronous arbiters.

Note that implementation wise they are different as the asynchronous ones are following a full-interlocked 4-phase return-to-zero protocol, whilst the synchronous set is following the synchronous elastic protocol~(SELF) which is a forward interlocked protocol~\cite{carmona2009elastic} but both inherit elastcity.

\end{itemize}

%\begin{figure}[!t]
%\centering
%\includegraphics[totalheight=5cm, width=0.45\textwidth]{Time-Borrowing}
%\caption{Elastic Buffers that govern elasticity on every link and the available time borrowing}
%\label{fig:Time-Borrowing}
%\end{figure}

%\subsection{Distributed Control}
%
%Unlike conventional control-driven architectures, in the eTeak synthesised circuits the control is distributed. It localises the synchronisation points between control unit and datapath which provide a way to interleave independent operations. This leads to an improved concurrency level and hence higher performance. In other words, operations are activated locally whenever data becomes available.  A distributed control/data architecture with relaxed global timing assumptions
%removes the concerns with the activation/compilation of the tasks when automatically refining the control flow by re-sizing the buffers.
%
%\subsection{Slack Elasticity}
%
%A \textit{Slack Elastic} system can be pipelined with any degree of storage on its communication channels.  This behaviour was first formalized for the distributed computation systems which were described in a CSP-like language, CHP. Slack elasticity provides a flexible communication environment for the computational blocks in the system.  This feature enables the tool to re-architecture the circuit through buffer insertion and buffer resizing which can influence the local/global cycle times and impact the critical path delays, consequently without affecting the functionality. 

\section{Problem Formulation}

In this section, we formulate the retiming approach in terms of area, power and performance in the context of dataflow architectures of eTeak. In this regard, basic CSP-based macromodules are exploited to describe our formulations. 

\subsection{Area Model} \label{intuitive_2}

In the dataflow circuits of eTeak, cell area is estimated using two main factors: the number of components {Join, Fork, Steer, Merge} and the Memory argument which includes a)~Storage where data (or instructions) are resident in the design, b)Variables which are essentially the primitives of the eTeak network and c)Buffers which are distributed across the network in form of FIFOs. The retiming technique proposed in this paper considers transforming, resizing, moving or inserting memories whilst preserving the number of components in the circuit.
 
\begin{equation}
Cell Area \propto {Comp + Mem}
\label{eq:area}
\end{equation}

\subsection{Power Model} \label{intuitive_1}

The dynamic and static power of our dataflow circuits can be modelled using the following equations:

\begin{equation}
Power_{dynamic} \propto {A\cdot f\cdot C\cdot V_{DD}^{2}}
\label{eq:powerd}
\end{equation}

where $(\frac{1}{2} \leq A \leq 1)$ as in our elastic controllers are latch based and hence level sensitive. $f, C,$ and $V$ are clock frequency, switched capacitance, and supply voltage, respectively. 

\begin{equation}
Power_{static} \propto {I_{leak}\cdot V_{DD}}
\label{eq:powerl}
\end{equation}

Note that $I_{leak}$ factor is proportional to the size of the circuit
which is represented as $Cell Area$. Therefore the larger the occupied silicon the higher $Power_{static}$ dissipation. Intuitively speaking, replacing the large \textit{always-on} memory blocks with smaller distributed memories across the circuit should contribute to the overall static power of the system.   

\begin{figure}[!t]
\centering
\includegraphics[totalheight=5cm, width=0.5\textwidth]{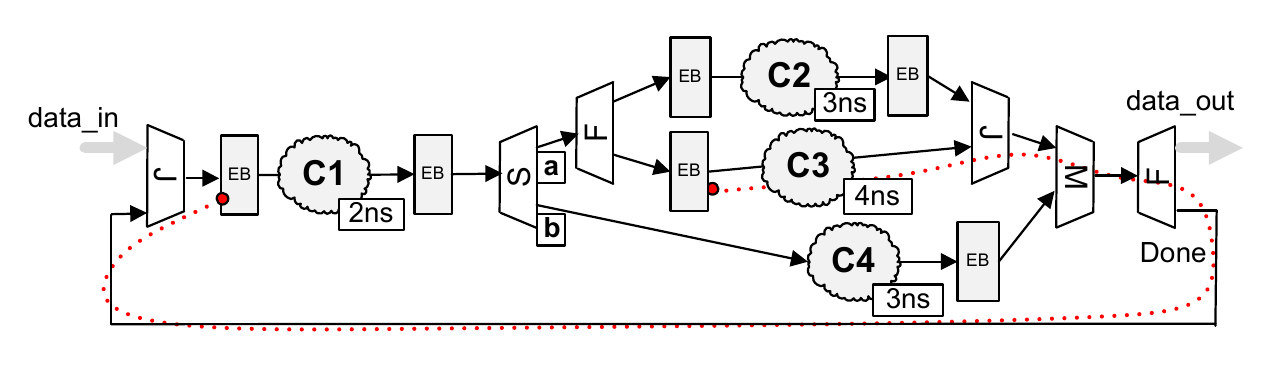}
\caption{an eTeak macro-module consist of primitive components. Data and control propagate together in a macro-module. The ${go/done}$ signals determine activation and termination.}
\label{fig:model}
\end{figure}

\subsection{Performance Model} \label{intuitive_3}

An intuitive understanding of the technique can be obtained by analysing the high-level macromodule model shown in Figure \ref{fig:model} which is a dataflow realisation of the expression C1 $;$ [(C2 \textbar\textbar C3) $?$ C4] in
the Balsa language; where C2 and C3 are specified as parallel computations and are
merged with C4. C1 is also considered in sequence with the rest. Each computation could represent a simple arithmetic unit or a complicated module, depending on the design's abstraction level. Blocks are annotated with arbitrary delays in this model for performance analysis.  The question mark in the expression implies
data-dependency and implies a Steer-Merge pair~(aka Choice) with $\alpha$ and $\beta$ probabilities on its $a$ and $b$ branches, respectively. The associated throughput is denoted by $\theta$ and is proportional to:

\begin{equation}
\theta \propto \dfrac{\sum m(e)}{\gamma\cdot\delta}
\label{eq:throughput}
\end{equation}

where $m$ denotes the sum of active tokens on the edges~(e) of the loop, which is assumed to be one in this particular example; $\delta$ $(= \frac{1}{f})$ represents the critical path delay which, in this example, equals to the associated delay of C3~(4~ns) plus the delay of the combinatorial components, which are assumed to be 0.1ns each.  Therefore the overall delay is 4.4~ns (the red dotted path). 

%\begin{figure}[!t]
%\centering
%\includegraphics[totalheight=5cm, width=0.45\textwidth]{motivate} 
%\caption{Variables vs. Channels: Poly's four various implementations which are transformable to each other via automatic retiming.}
%\label{fig:motivate}
%\end{figure}

\section{Methodology: Power/Performance Exploration}

\subsection{Motivational Example}

Memory distribution and orchestration in a dataflow circuit has considerable effects on area/power/performance. Moreover proper buffering is necessary to ensure deadlock and starvation freedom. The loop structures in the circuit require enough buffering for a lead token to move forward and leave space for the next one. In the eTeak synthesis flow, no buffering is introduced initially. This helps the designers to explore the different buffering policies.

In this regard, the most simple strategy is to buffer every link. In our 1-safe models this guarantees deadlock freedom but the overhead would be dominant. Alternative approach is to map networks onto a directed graph without caring about the component types, then a depth-first-search technique is exploited to identify the back-edges to indicate cycles~(loops). Buffers are inserted in the proceeding and succeeding links. Based on the topology of the circuit, some extra buffers may be added. This approach has no insight of the circuit and performs the same strategy. We have discovered that by having a high-level notion of the description, a more efficient strategy could be employed.

Figure~\ref{fig:balsa} shows a Balsa implementation of a Polynomial evaluation algorithm wherein the \textit{While loop} (lines 6 -- 16) iterates for \textit{degree} times. Within the loop the \textit{for construct} keeps receiving coefficients via channels (three in this example) and writes out the calculated value into \textit{addRes} channel.\\

%\colorbox{pink}{Figure ~\ref{fig:motivate} depicts} four different dataflow implementations of the high-level description which are transformable to each other via retiming. Our main interest in this work is to identify which strategy is suitable for a given application with a set of constraints. (a)~shows three adders with the associated channels that are writing into a single variable. To ensure correctness write-after-read links are buffered to avoid deadlock. (b)~shows by breaking the variable in (a) into several single assignment variables (aka buffers) the loops write-after-read cycles are removed. (c)~shows a single channel implementation of the Poly example in which time multiplexing is required which needs extra buffering to ensure that the data generated from the previous stages is not lost. Finally (d) shows a pipelined approach that the adders are chained via channels. In this case the depth of the channels represents the level of \colorbox{pink}{pipelining.}    

The next section describes our proposed automatic approach.

\begin{figure}
\begin{algorithmic}[1]
\Procedure{POLY}{$x,coefs[0..degree],a$}\Comment{ inputs: x, coefs[0..deg],  output: c}
\State \textbf{loop}
\State $xV \gets x$ \textbf{\textbar \textbar}
\State $aV \gets coefs[deg]$
\State $i =  deg - 1 $ \textbf{;}
\While{$i[31] \not=1$}\Comment {i \textgreater =0}
\State  mul($xV->,aV->,temp$)
\State  \textbf{case} $i$ of
\For{\texttt{j in 0..degree}}
\State $temp , coefs[j]$ --\textgreater then
\State $addRes\gets temp + coefs[j] $
\State end
\EndFor \textbf{ \textbar \textbar}
\State $aV = addRes $
\State $i =  i - 1 $
\EndWhile\label{euclidendwhile}
\State \textbf{;} $a \gets aV$ 
\State \textbf{end loop}
\EndProcedure
\end{algorithmic}
\caption{Balsa Model of a Polynomial Evaluation}
\label {fig:balsa}
\end{figure}

%\begin{figure}[!t]
%\centering
%\includegraphics[width=3.5 in] {MyAlgPoly.pdf}
%\caption{The resulting buffered Polynomial example synthesised using eTeak.}
%\label{fig:buffered}
%\end{figure}

Our automatic approach consists of two phases: 1) extracting the design dependent constraints that should be considered to preserve functionality (extracting the dependency graph) which is explained in this section, and 2) Deploying buffers which will be discussed in the next section.

\subsection{Basic Graphical Model}

Initially the high level spec of a design expressed in the Balsa language is analysed and a dependency graph is generated.

\begin{definition}
The Dependency Graph is a disjoint directed graph \textit{G(V,E)} where the source
nodes \textit{Vs} consist of variables or channels and the destination nodes \textit{Vd} determines variables, channels or components. The edge \textit{(u,v)} represents a WAR or a PAC constraint on \textit{u} in which \textit{u} is read or
consumed by \textit{v}.
\end{definition}
 The edges in the graph represent the following constraints: 
\subsubsection {WAR and RAW Constraints}

On every shared memory (aka variable) write-after-read~(WAR) and read-after-write~(RAW) constraints are considered to ensure that the value in the variable is not over written or read after the write action takes place. WAR and RAW constraints should get buffered properly as they might raise deadlocks.

\subsubsection {PAC Constraints}

In the CSP model procedures communicate through channels. Due to the slack elastic nature of the graphs the channels may have bounded degree of storage~(0 to n). In CSP variables are defined within procedures. By decomposing the procedures into producer and consumer sets PAC~(Produce After Consume) has to be strictly followed: The consumer should consume the data only after the producer has it produced. Also when there is a backward loop in the dataflow, the producer should produce the next data only after the consumer has consumed it.

The beautiful fact about PAC is that it is able to view the design in form of producers and consumers at a fine level of granularity. This implicitly transforms every variable within the procedure to a buffer on a communication channel. Since it is possible to see every storage unit (e.g. DRAM) as a procedure with input and output access channels PAC could simply replace a DRAM with several distributed shared memory units (e.g. caches) across the network. The same is applicable to transform shared memory units into several communication memories (e.g. FIFOs) which are usually allocated on on-chip fast BRAMs on FPGAs. 

\subsection{Deploying Buffers}

By having the information from the dependency graph and the eTeak `not-buffered' network of components, two phases are considered for this part of the algorithm: 1) Mark the potential links for buffering (extracted from the dependency graph) 2) Retime the buffered links based on the next component and buffer the desired links. Note that our approach is optimized to avoid inserting multiple buffers in the same link.

\subsubsection{Phase1: Marking the potential links for buffering}
For each dependency in the graph, based on the source and destination nodes, a different marking policy is used. The source nodes are channels or variables and destination nodes could vary as explained above. In case of a variable, the first link to be marked is the link after the Read portion of it.  For channels, it is the link after associated with its consumer. The next link to be marked is based on the destination node type. As a whole rule it could be considered as the link after the production which its definition varies in different types. 
\iffalse
\begin{itemize}
\item{Variable to Variable} 
 a) Mark the link after the Read Portion of the source node (ReadDone) and b) The link after the write portion of the destination node (WriteDone)
\item {Variable to Operation} 
a) Mark the link after the Read Portion of the source node (ReadDone) and b) The link after the operation
\item {Variable to Other components} 
a) Mark the link after the Read Portion of the source node (ReadDone) and b) The writeGo link of the variable
\item {Variable to channel} 
a) Mark the link after the Read Portion of the source node (ReadDone) and b) The link after the channel's production
\item {Channel to Variable} 
a) Mark the link after the channel's consumption and b) The link after the variable's write portion (WriteDone)
\item {Channel to another module} 
a) Mark the link after the channel's consumption and b) Mark end of the module (Done signal and outputs)
\item {Channel to Channel} 
a) Mark the link after the source's consumption and b)The link after the destination's production
\item {Channel to operation} 
a) Mark the link after the source's consumption and b) The link after the operation.\\
\end{itemize}
\fi

\subsubsection{Phase 2: retiming and buffering based on the marked links}
In this phase, each marked link is taken and the next component in the network is checked, based on the upcoming component the strategy is different. In case of a join, two choices of buffering could be considered, buffering all the inputs~(including the marked link) or retime it to the output which is considered in our algorithm. The other components are considered either based on their unique properties and the impacts on area/power/performance.

For the top level networks without a go signal, the links before and after the I component should be buffered as well.  This makes sure that the initialization cycle that helps the circuit start over has enough buffering for the tokens to progress. In Synchronous Elastic circuits (not asynchronous) extra buffering of merges/steers may be needed for balancing the data and control. \\

%Consider the POLY example. By having the dependency graph shown in Figure \ref{fig:dep1} and the eTeak network of components, the algorithm is 
%applicable.  Figure \ref{fig:buffered} illustrates the resulting buffered network.

\begin{figure}[!t]
\centering

\begin{subfigure}[t]{0.4\textwidth}
        \includegraphics[width=\textwidth]{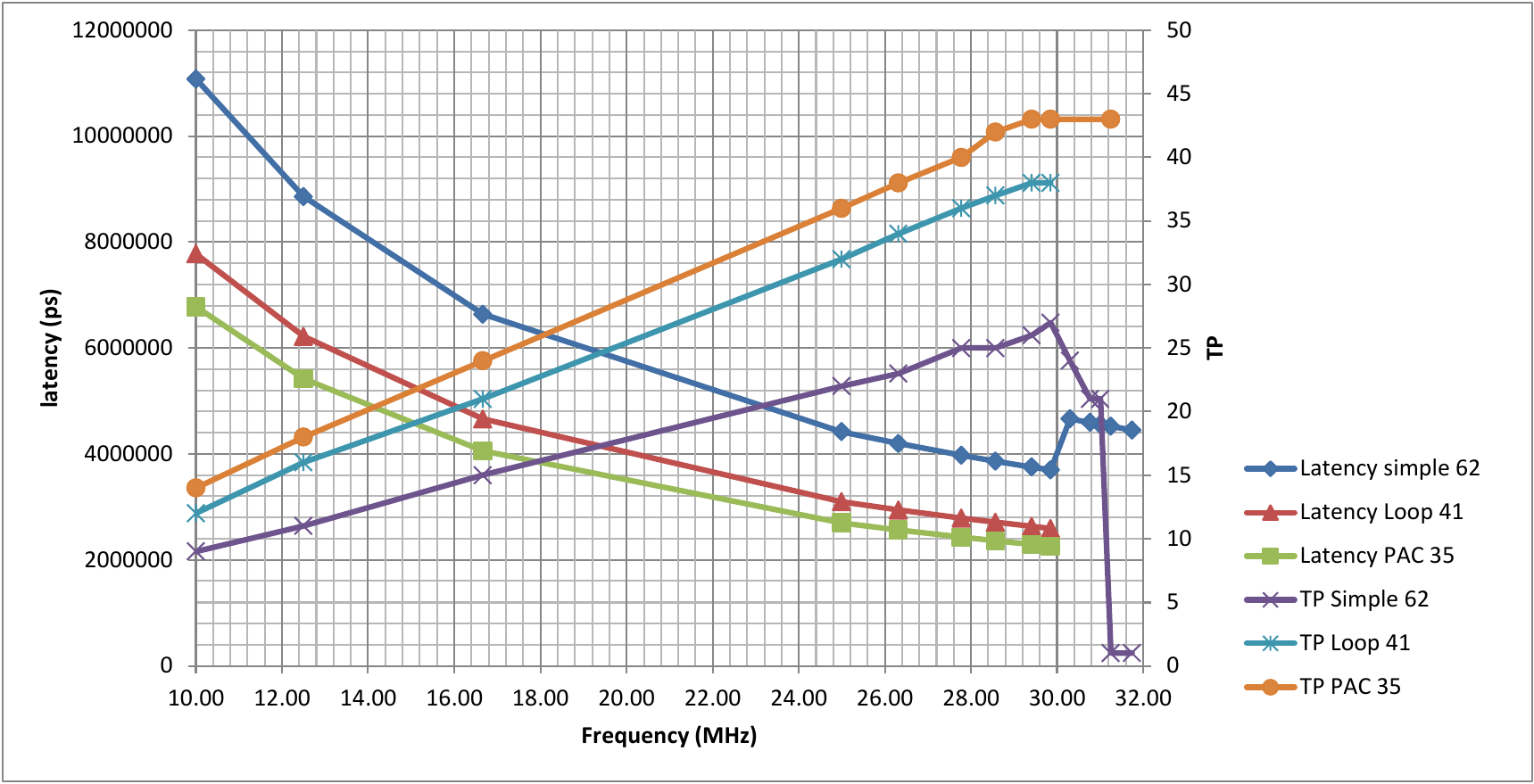}
        \caption{Latency and throughput comparison}
        \label{fig:ElgLa1}
    \end{subfigure}
   % ~ %add desired spacing between images, e. g. ~, \quad, \qquad, \hfill etc. 
%    \begin{subfigure}[t]{0.4\textwidth}
%        \includegraphics[width=\textwidth]{2}
%        \caption{Throughput-The Number of outputs in a same fixed time }
%        \label{fig:ElgTp}
%    \end{subfigure}
    %~ %add desired spacing between images, e. g. ~, \quad, \qquad, \hfill etc. 
%    \begin{subfigure}[t]{0.4\textwidth}
%        \includegraphics[width=\textwidth]{3}
%        \caption{Latency comparison for the other input set}
%        \label{fig:ElgLa2}
%    \end{subfigure}
% ~ %add desired spacing between images, e. g. ~, \quad, \qquad, \hfill etc. 
    \begin{subfigure}[t]{0.4\textwidth}
        \includegraphics[width=\textwidth]{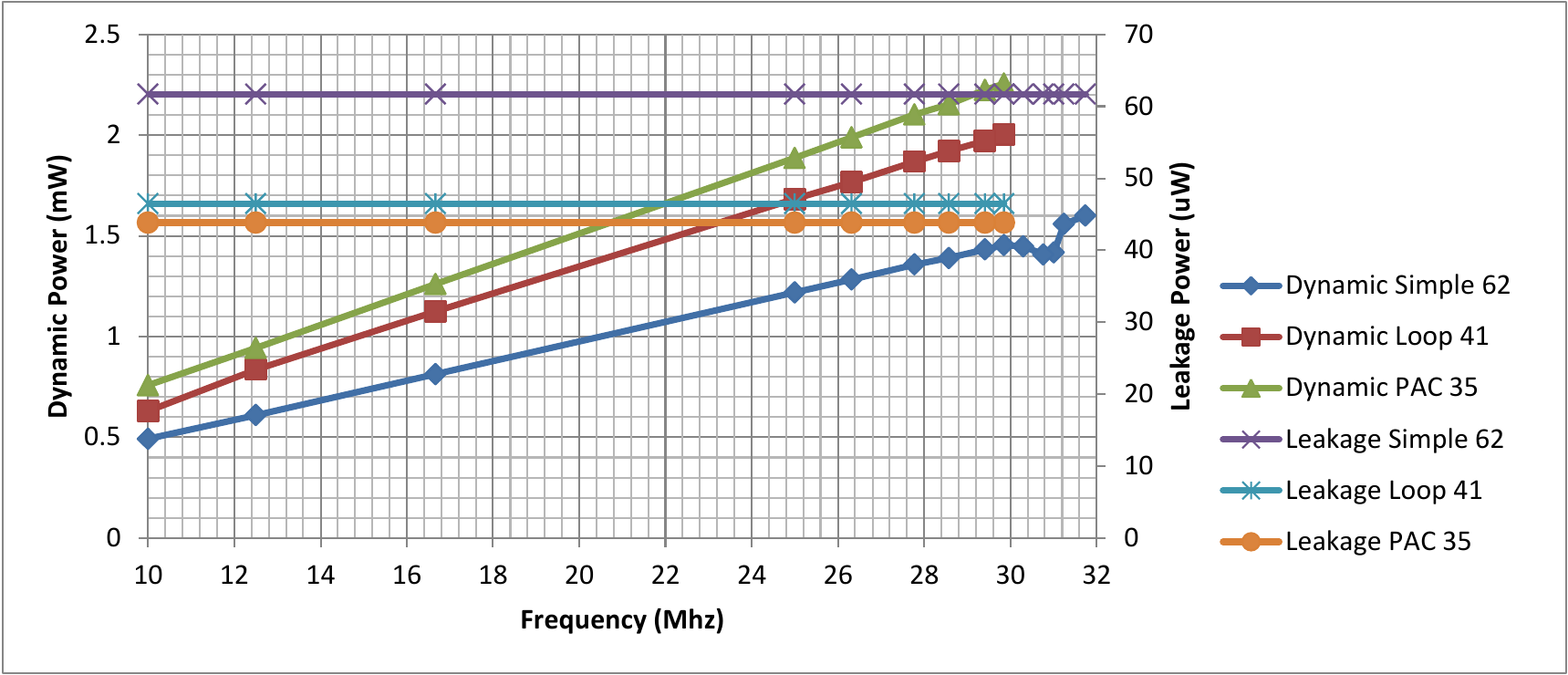}
        \caption{Dynamic and Leakage Power Comparison}
        \label{fig:ElgDy}
    \end{subfigure}
    %~ %add desired spacing between images, e. g. ~, \quad, \qquad, \hfill etc. 
%    \begin{subfigure}[t]{0.4\textwidth}
%        \includegraphics[width=\textwidth]{5}
%        \caption{ Static Power (Leakage Power) Comparison}
%        \label{fig:ElgSp}
%    \end{subfigure}

\caption{The comparison results for the approaches in Synchronous Elastic ElGcd benchmark }
\label{fig: elgRes}
\end{figure}

\begin{figure}
\centering

\begin{subfigure}[t]{0.4\textwidth}
        \includegraphics[width=\textwidth]{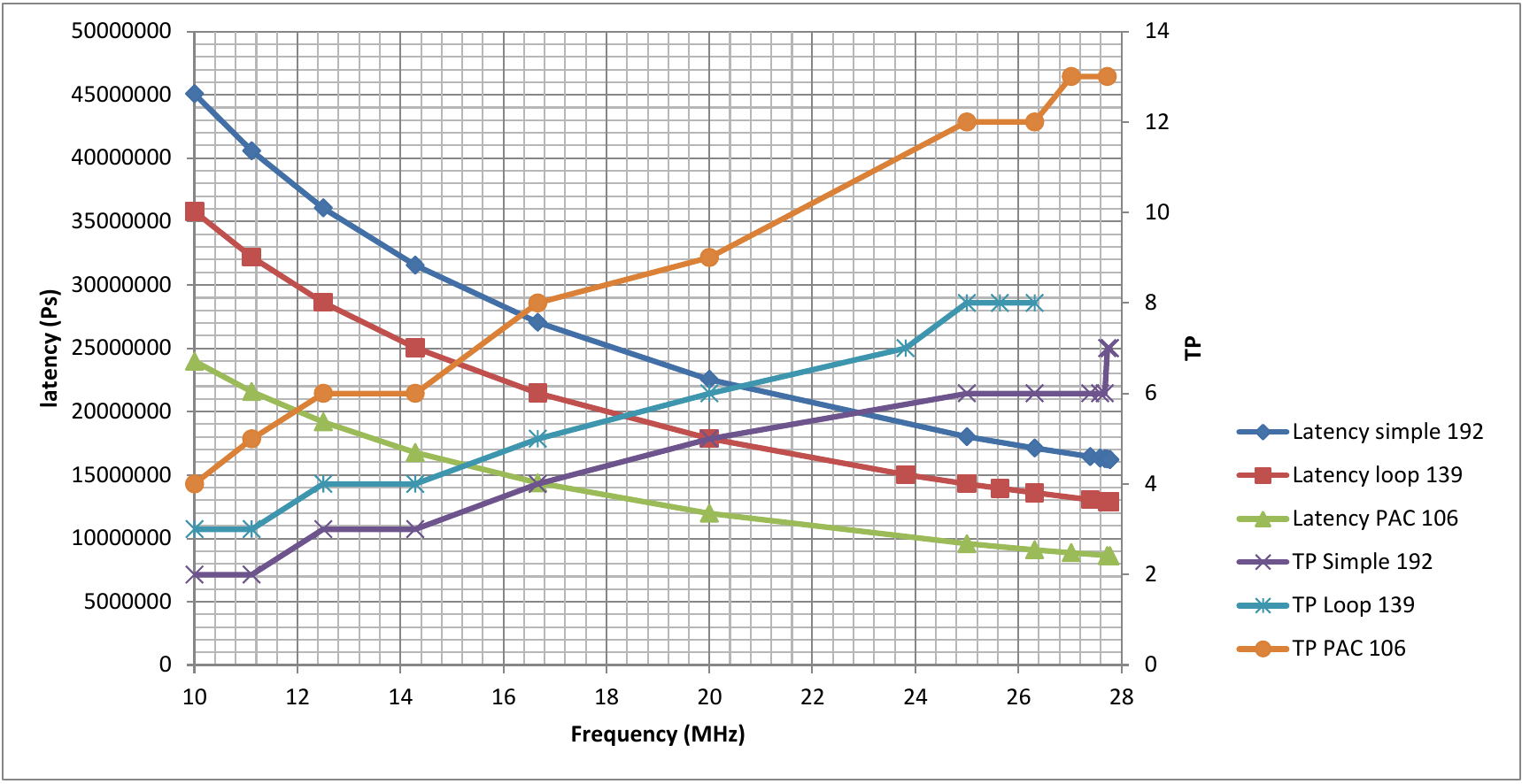}
        \caption{Latency and throughput comparison}
        \label{fig:PolLa1}
    \end{subfigure}
   % ~ %add desired spacing between images, e. g. ~, \quad, \qquad, \hfill etc. 
%    \begin{subfigure}[t]{0.4\textwidth}
%        \includegraphics[width=\textwidth]{polyTp}
%        \caption{Throughput-The Number of outputs in a same fixed time }
%        \label{fig:PolTp}
%    \end{subfigure}
    %~ %add desired spacing between images, e. g. ~, \quad, \qquad, \hfill etc. 
%    \begin{subfigure}[t]{0.4\textwidth}
%        \includegraphics[width=\textwidth]{polyLatency2}
%        \caption{Latency comparison for the other input set}
%        \label{fig:PolLa2}
%    \end{subfigure}
% ~ %add desired spacing between images, e. g. ~, \quad, \qquad, \hfill etc. 
    \begin{subfigure}[t]{0.4\textwidth}
        \includegraphics[width=\textwidth]{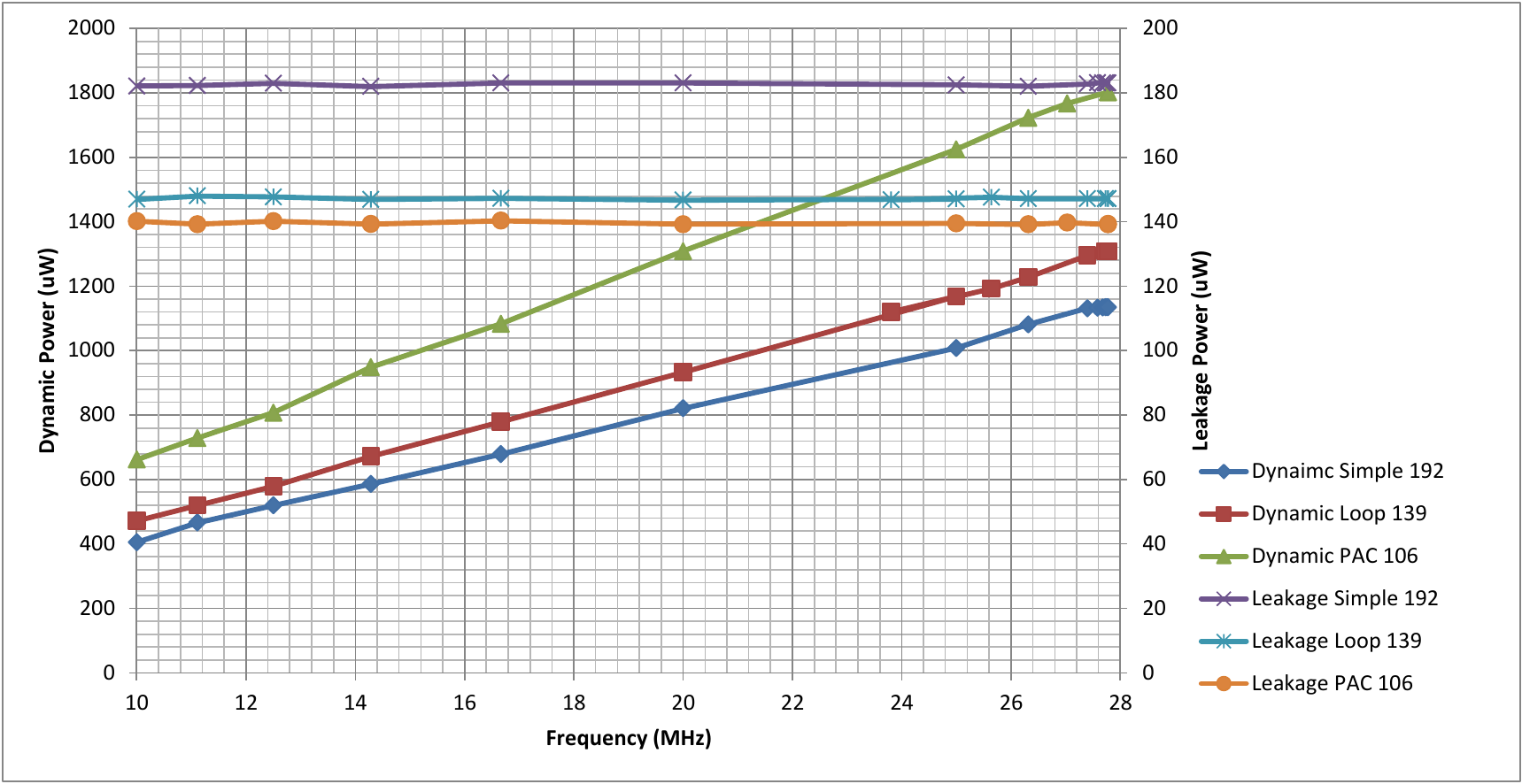}
        \caption{Dynamic and Leakage Power Comparison}
        \label{fig:PolDy}
    \end{subfigure}
    %~ %add desired spacing between images, e. g. ~, \quad, \qquad, \hfill etc. 
%    \begin{subfigure}[t]{0.4\textwidth}
%        \includegraphics[width=\textwidth]{polyLeakage}
%        \caption{ Static Power (Leakage Power) Comparison}
%        \label{fig:PolSp}
%    \end{subfigure}

\caption{The comparison results for the approaches in Synchronous Elastic POLY benchmark }
\label{fig:poly}

\end{figure}

%%SMul
\begin{figure}
\centering

\begin{subfigure}[t]{0.4\textwidth}
        \includegraphics[width=\textwidth]{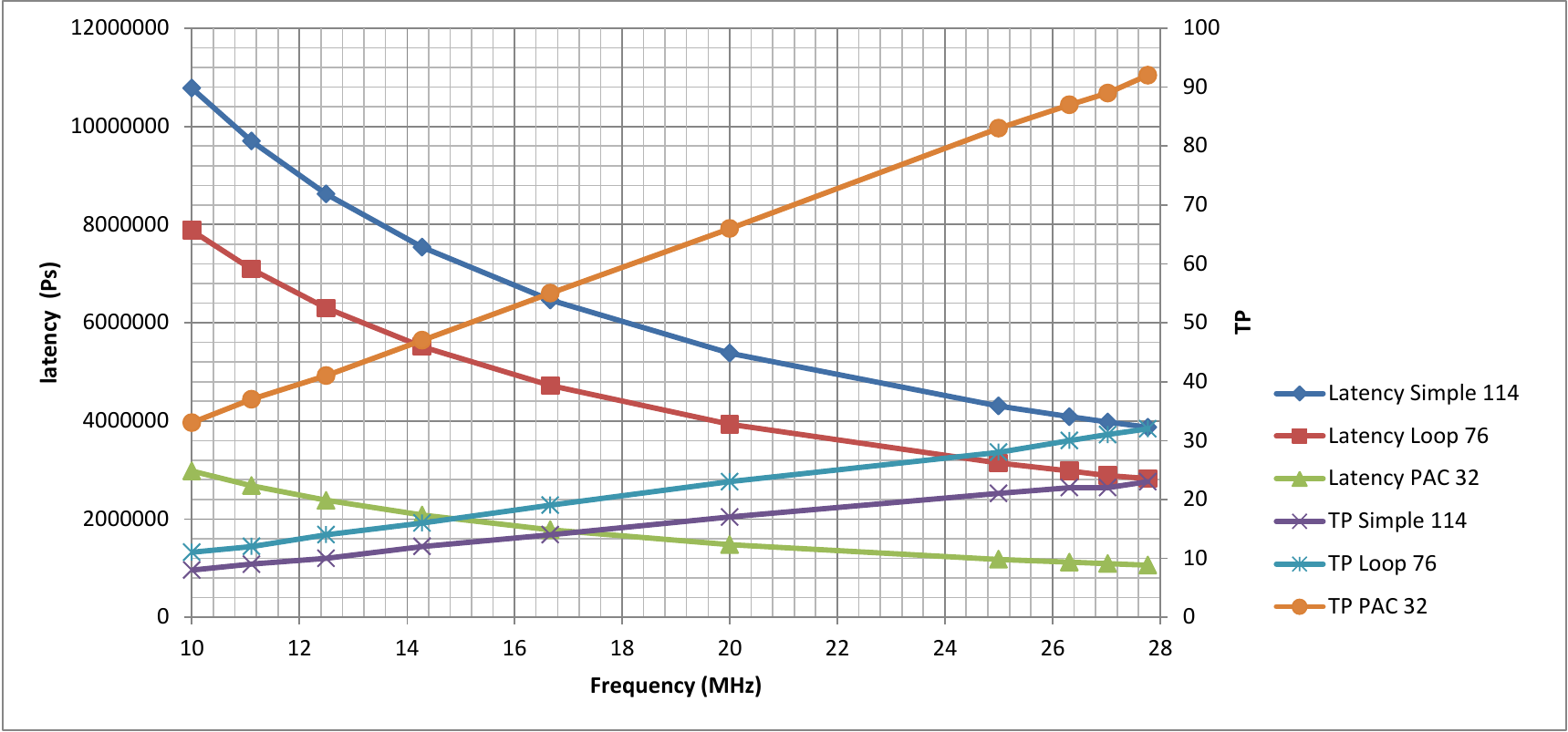}
        \caption{Latency and throughput comparison}
        \label{fig:mulLa1}
    \end{subfigure}
   % ~ %add desired spacing between images, e. g. ~, \quad, \qquad, \hfill etc. 
%    \begin{subfigure}[t]{0.4\textwidth}
%        \includegraphics[width=\textwidth]{smulTp}
%        \caption{Throughput-The Number of outputs in a same fixed time }
%        \label{fig:mulTp}
%    \end{subfigure}
    %~ %add desired spacing between images, e. g. ~, \quad, \qquad, \hfill etc. 
%    \begin{subfigure}[t]{0.4\textwidth}
%        \includegraphics[width=\textwidth]{smulLatency2}
%        \caption{Latency comparison for the other input set}
%        \label{fig:mulLa2}
%    \end{subfigure}
% ~ %add desired spacing between images, e. g. ~, \quad, \qquad, \hfill etc. 
    \begin{subfigure}[t]{0.4\textwidth}
        \includegraphics[width=\textwidth]{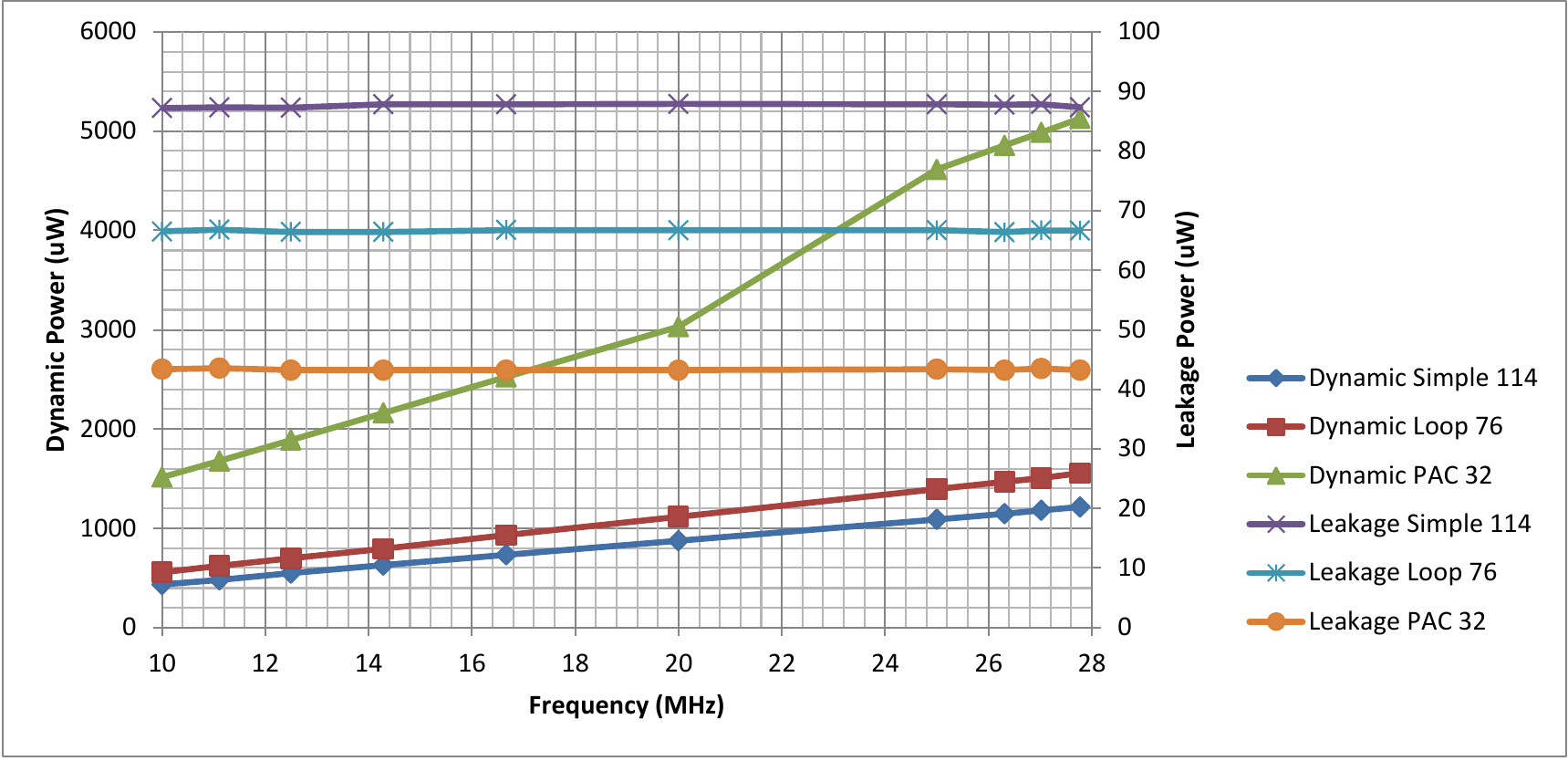}
        \caption{Dynamic and Leakage Power Comparison}
        \label{fig:mulDy}
    \end{subfigure}
    %~ %add desired spacing between images, e. g. ~, \quad, \qquad, \hfill etc. 
%    \begin{subfigure}[t]{0.4\textwidth}
%        \includegraphics[width=\textwidth]{smulLeakage}
%        \caption{ Static Power (Leakage Power) Comparison}
%        \label{fig:mulSp}
%    \end{subfigure}

\caption{The comparison results for the approaches in Synchronous Elastic SMUL benchmark }
\label{fig:smul}

\vspace{-0.7cm}
\end{figure}

\section {Experimental Results}

\subsection {Benchmarks}   % needs a review
 Our approach targets accelerators especially iterative algorithms with data dependent loops and lots of data dependencies. We tested implementations of the followings as benchmarks:
\subsubsection {ELGCD} The Euclid’s GCD solver which derives the greatest common divisor of the inputs.
\subsubsection {POLY} The Polynomial evaluation algorithm which can be executed with arbitrary degree of coefficients.
%\subsubsection{DDLC} It is a test algorithm with a data-dependent loop that performs data-dependent computations.
%\subsubsection{SORTER} A bitonic sorter.
\subsubsection{SMUL} A set of signed multiplications on signed inputs.
%\subsubsection{PAC} A producer/consumer model in which data is produced and consumed several times in a loop.

\begin{table*}[!t]
\caption{The metrics associated with every application for both Synchronous Elastic and Asynchronous (pure Elastic) implementations}
\centering
\begin{tabular}{| l | l | l | l | l | l | l |}
\hline
	Application/~&~No.~of&No.~of&No.~of& Buffer~Reduction& No.~of& Buffer Reduction\\ 
	Approach~&~Teak~Comps~&~links&Buffers/SE&(SE)\% &Buffers/AS&(AS)\% \\ \hline	
	ELGCD/Simple & 38 & 62 & 62 & - & 62 & - \\ \hline
	ELGCD/loop & 38 & 62 & 41 & 33.87 \% & 36 & 41.93 \%\\ \hline
	ELGCD/PAC & 38 & 62 & 35 & 43.54 \%& 11 & 82.26 \% \\ \hline
	POLY/Simple & 121 & 192 & 192 & - & 192 & - \\ \hline
	POLY/loop & 121 & 192 & 139 & 27.60 \%& 104 & 45.83 \%\\ \hline
	POLY/PAC & 121 & 192 & 106 & 44.79 \% & 32 & 83.33 \%\\ \hline
	SMUL/Simple & 70 & 114 & 114 & - & 114 & - \\ \hline
	SMUL/loop & 70 & 114 & 76 & 33.33 \%& 74 & 35.09 \%\\ \hline
	SMUL/PAC & 70 & 114 & 32 & 71.93 \%& 16 & 85.96 \% \\ \hline
	%%SORTER/Simple & x & x & x & x & x & x \\ \hline
	%%SORTER/loop & x & x & x & x & x & x \\ \hline
	%%SORTER/PAC & x & x & x & x & x & x \\ \hline
\end{tabular}
\label {app}
\end{table*}

 \begin{table*}[!t]
\caption{The Experimental Results for Asynchronous (Pure Elastic) implementations}
\centering
\begin{tabular}{|l | l | l | l | l | l | l|}
\hline
	Benchmark & No.~of& Throughput & Latency1 & Latency2 & DynamicPower & LeakagePower \\
	&Buffers& MBit/s & (ps) & (ps) & (uw) & (uw) \\ \hline
	ElGcd/simple & 62 & 114 & 881000 & 1729000 & 12.1689 & 32.2589 \\ \hline
	ElGcd/loop & 36 & 140 & 705000 & 1369000 & 14.6304 & 25.8753 \\ \hline
	ElGcd/PAC & 11 & 152 & 634000 & 1225000 & 13.939 & 21.3084 \\ \hline
	SMUL/simple & 114 & 132 & 675000 & 1499000 & 12.5316 & 50.5335 \\ \hline
	SMUL/loop & 74 & 144 & 616000 & 1361000 & 13.3248 & 43.0839 \\ \hline
	SMUL/PAC & 16 & 129 & 671000 & 1318000 & 10.4344 & 35.8966 \\ \hline
	POLY/simple & 192 & 39 & 3047000 & 5078000 & 12.5376 & 100.0439 \\ \hline
	POLY/loop & 104 & 41 & 2795000 & 4802000 & 12.2293 & 73.5797 \\ \hline
	POLY/PAC & 32 & 40 & 2914000 & 4762000 & 12.0182 & 71.8703 \\ \hline
\end{tabular}
\label {perft}
\end{table*}

\subsection {Experimental Setup}

The benchmarks are implemented in the Balsa language and synthesised using eTeak. Both \textit{Asynchronous} and \textit{Synchronous Elastic} implementations of the benchmarks are studied. For each benchmark three different buffering policies are considered: \textit{(i)~Simple policy}: It buffers all the links in the dataflow, which gains the shortest local cycle time in the asynchronous designs and, of course, shortest critical path in the synchronous domain. This policy is exhaustive and imposes significant overhead. \textit{(ii)~loop policy}: the classic asynchronous method where every loop in the design is buffered \textit{blindly} to avoid deadlocks. \textit{(iii)~PAC policy}: The proposed methodology in this work.\\
Every design is technology mapped using eTeak's rich backend. The 90nm technology node is used in our experiments. All of the datapath components are re-synthesised and tech mapped using Synopsys' Design~Compiler and used in the eTeak networks; Two different data sets are fed into the designs for a certain amount of time. Note that the datapath components (like adders) are combinatorial logic and impose a long critical path. However this dose not affect our retiming methodology as we focus on channles and variables. \\
For asynchronous circuits, the throughput is determined by measuring the average cycle time of the circuit. The cycle time is the delay between producing consecutive results in the repeated operation. Throughput is defined as the number of results spitted out in a certain period of time. Both the asynchronous and synchronous elastic dataflow circuits show a data-dependent behaviour with different input data sets. In the synchronous elastic circuits `clock' is introduced which unlike the asynchronous designs is defined based on the worst-case cycle time (aka critical path). A range of clock frequencies has been tested for synchronous elastic circuits. 
The buffering policies vary the number of buffers on WAR/RAW/PAC links whilst the overall structure remains almost the same. Therefore, reducing the number of buffers has direct impact on area.

\vspace{-0.5cm}

\subsection {Results and Discussion} 

Table~\ref{app} depicts the metrics associated with each application. Number of teak components is the same in different implementations of each benchmark. However the implementation details of the components change when moving from the asynchronous to the synchronous elastic domain as the protocol changes from 4-phase return-to-zero to a forward-interlocked protocol. The buffer reduction rates compared to Simple policy are illustrated in the table. It could be observed that the PAC methodology reduces the number of buffers up to nearly 72\% in synchronous elastic and up to nearly 86\% in asynchronous circuits compared to simple policy. The improvement against Loop is 57\% and 78\% in synchronous elastic and asynchronous circuits, respectively.

Table~\ref{perft} presents the results in the asynchronous domain. The leakage power is reduced by 35\% which is the impact of decreasing the size of the circuit. This has consequences on throughput which is expectable as the asynchronous circuits benefit from over buffering as it shortens the cycle time in average. The second input data set has higher latency. Therefore the latency is reduced compared to both loop and simple policies. A slight decrease in POLY and SMUL is observed which is negligible.

Figure~\ref{fig: elgRes} demonstrates the combo results associated with the synchronous elastic ElGcd. (a)~shows the impact of the clock frequency~($\delta$) in different buffering approaches where the cycle time~($\gamma$) varies in different buffering policies. It can be seen that the throughput ($\theta$ = Tp) is higher in PAC (figure \ref{fig:ElgLa1}-throughput) compared to its Loop and Simple counterparts. Note that in the synchronous elastic circuits with Simple buffering over clocking the circuit results in lower latency (figure~\ref{fig:ElgLa1}-latency) however the produced results are \textit{approximately} correct as the critical path delay is violated. This does not affect functionality as long as the control bits enclosed in data are safe.    

In general PAC results in lower latency. By reducing the number of elastic buffers the overall latency decreases. The leakage power in Figure~\ref{fig:ElgDy}-Dynamic which is proportional to area is decreased while reducing the number of buffers across the selected policies. Dynamic power in Figure~\ref{fig:ElgDy}-Leakage increases linearly as the clock frequency increases. The penalty of increasing throughput and decreasing the leakage power in PAC is reasonable when considering the performance/power factor against the other policies. 

In POLY~(figures~\ref{fig:poly}) and MUL~(Figure~\ref{fig:smul}) the PAC impact on performance is at least twice higher than GCD. PAC attempts to decompose the shared memories (variables) and replace them with buffers (communication memories). Although in all of the considered descriptions the number of variables are equal, their utilisation by the explicit control varies. For instance, in GCD variables are within a \textit{While-loop} which exerts limitation on the decomposition performed by PAC. Whilst in POLY and MUL variables are regularly distributed and accesses are less control dominant which makes PAC more effective.

%\vspace{-0.25cm}
%\section{Related Work}
%
%Elastic circuits have been around for a long time and have shown significant advantages over their rigid synchronous counterparts. We were particularly interested in the concept of slack elasticity in the CSP-based class of circuits. Cortadella et. al~\cite{bufistov2008performance} have shown that retiming and recycling can be employed to refine the pipelining depth. Although their processor case study enjoys fine-grained elasticity, their control driven computation model with global control signalling expressed at RTL limits the set of transformations toward automatic pipelining. CSP model scales well and this has been the reason why Google is exploiting it in cloud programming at software level. There are many works including CASPAR~\cite{Gangwani2016Caspar} that have attempted to lower the overhead of synchronisation by pushing the task of barrier insertion to the developers but their scaliblity is still an issue unless they exploit distributed queues at hardware level which again borrows the concept from CSP. Although elasticity has major benefits in realising adaptivity, its presence at fine granularity level appears to be prohibitive.  

\vspace{-0.25cm}

\section{Conclusion}

We introduced an automatic retiming technique which refines the CSP based dataflow accelerators of eTeak via buffer resizing/insertion to meet application's power, area or performance constraints. Our experiments demonstrate how different memory arrangements across the design could impact the clock frequency and, consequently, the throughput and power consumption.

The slack elastic nature of the CSP dataflows allows us to explore different degrees of pipelining without affecting the functionality. By separating functionality from timing we demonstrate that most of the compiler-level transformations are applicable at later phases where timing takes place.

As future work, we exploit this technique to investigate architectural scaling in the FPGA domain where compile and synthesis takes so long. It is extremely necessary to leverage flexible architectures that satisfy constraints with minimal structural changes such as varying the FIFO depths or changing the degree of pipelining.

\section{Acknowledgement} \label{achnowledge}

We would like to thank reviewers at Advanced Processor Technologies of The University of Manchester and Digital Systems Design Lab of The University of Tehran for their valuable feedback. Mahdi holds EPSRC Doctoral Prize Fellowship 2015-17 at The University of Manchester. 

\bibliographystyle{IEEEtran}
%\bibliography{references}

\vspace{10cm}

\begin{IEEEbiography}[{\includegraphics[width=1in,height=1.25in,clip,keepaspectratio]{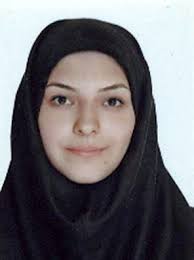}}]{Ana Lava}
received the BSc degree from the Shahrood University and MSc degree from the University of Tehran, Iran, in 2013 and 2016, respectively, all in computer engineering. She is currently pursing a PhD research position in the area of high level synthesis.
\end{IEEEbiography}

\vspace{-10cm}

\begin{IEEEbiography}[{\includegraphics[width=1in,height=1.25in,clip,keepaspectratio]{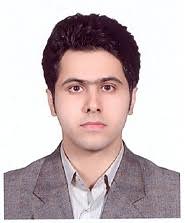}}]{Mahdi Jelodari Mamaghani}
received the BSc degree in computer engineering from the University of Tehran, Iran, in 2012, and PhD degree in computer science in 2016 from the University of Manchester, UK. Currently he is holding an EPSRC doctoral prize fellwoship at the University of Manchester, UK. In 2015 his contribution on high level synthesis of elasticity won the Best DATE IP award. He is also an IEEE member since 2010. 
\end{IEEEbiography}

\vspace{-10cm}

\begin{IEEEbiography}[{\includegraphics[width=1in,height=1.25in,clip,keepaspectratio]{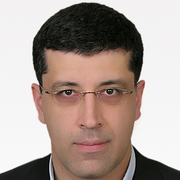}}]{Siamak Mohammadi}
received the BSc, MSc, and PhD degrees from the University of Paris Sud Orsay, France, in 1990, 1992, and 1996, respectively, all in electrical engineering. From 1997 to 1999, he was a research associate with the Department of Computer Science, University of Manchester, England. In 1999, he moved to Canada and worked in industry until 2005. He is currently an assistant professor in the School of Electrical and Computer Engineering, at the University of Tehran, Iran.
\end{IEEEbiography}
\vspace{-10cm}

\begin{IEEEbiography}[{\includegraphics[width=1in,height=1.25in,clip,keepaspectratio]{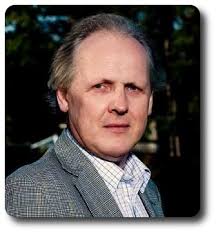}}]{Steve B. Furber}
(M’98-SM’02-F’05) is an ICL professor of computer engineering in the School of Computer Science at the University of Manchester. He was at Acorn Computers during the 1980s, where he led the development of the first ARM microprocessors. He is a fellow of the Royal Society, the Royal Academy of Engineering, the British Computer Society, the Institution of Engineering and Technology, and the IEEE.
\end{IEEEbiography}

\end{document}